\documentclass[proceedings, preprint]{rmaa}

\usepackage{paralist}

\usepackage{psfrag,color}

\SetYear{2019}
\SetConfTitle{Workshop on Robotic Autonomous Observatories (AstroRob 2019)}

\title{The 4 meter New Robotic Telescope project: an updated report} 

\author{
  C.~M. Guti\'errez,\altaffilmark{1,2} 
  M. Torres,\altaffilmark{1,2}
  A. Oria,\altaffilmark{1,2}
  J.~J. Fern\'andez-Valdivia,\altaffilmark{1,2}
  D. Arnold,\altaffilmark{3}
  D. Copley, \altaffilmark{3}
  C. Copperwheat,\altaffilmark{3}
  J. de Cos Juez,\altaffilmark{4}
  A. Franco,\altaffilmark{1,2}
  Y. Fan,\altaffilmark{5} 
  A. Garc\'\i a Pi\~nero,\altaffilmark{1,2}
  E. Harvey,\altaffilmark{3}
  H. Jermak,\altaffilmark{3}
  X. Jiang,\altaffilmark{6} 
  J.~H. Knapen,\altaffilmark{1,2}
  A. McGrath,\altaffilmark{3}
  A. Ranjbar,\altaffilmark{3}
  R. Rebolo,\altaffilmark{1,2,7}
  R. Smith,\altaffilmark{3}
  I.~A. Steele,\altaffilmark{3}
  Z. Wang,\altaffilmark{8} 
  X. Wu,\altaffilmark{9} 
  D. Xu,\altaffilmark{10} 
  S. Xue,\altaffilmark{10} 
  W. Yuan,\altaffilmark{10} 
 and Y. Zheng\altaffilmark{11}  
}

\altaffiltext{1}{Instituto de Astrof\'\i sica de Canarias, E-38205
Tenerife, Spain}

\altaffiltext{2}{University of La Laguna, E-38206 Tenerife, Spain}

\altaffiltext{3}{Astrophysics Research Institute, Liverpool John Moores University,
IC2, Liverpool Science Park,
146 Brownlow Hill, Liverpool, L3 5RF,UK}

\altaffiltext{4}{Instituto Universitario de Ciencias y Tecnolog\'\i as Espaciales de Asturias
E-33004, C/Independencia 13-Oviedo, Spain}

\altaffiltext{5}{Yunnan Astronomical Observatories, Chinese Academy of Sciences, Kunming 650011, China}

\altaffiltext{6}{National Astronomical Observatories, Chinese Academy of Sciences, Beijing 100101, China}

\altaffiltext{7}{Consejo Superior de Investigaciones Cient\'\i ficas, E-28006 Madrid, Spain}

\altaffiltext{8}{Shanghai Astronomical Observatory, Chinese Academy of Sciences, Shanghai 200030, China}

\altaffiltext{9}{Purple Mountain Observatory, Chinese Academy of Sciences, Nanjing 210008, China}

\altaffiltext{10}{CAS Key Laboratory of Space Astronomy and Technology, National Astronomical Observatories, Chinese Academy of Sciences, Beijing 100101, China}

\altaffiltext{11}{National Astronomical Observatories, Nanjing Institute of Astronomical Optics and Technology (NIAOT), Chinese Academy of Science, 188 Bancang Street, Nanjing 210042, China}

\shortauthor{Guti\'errez et al.}
\shorttitle{The 4 meter New Robotic Telescope}

\listofauthors{W. J. Henney, A. Collaborator, \& L. Author}
\indexauthor{Henney, W. J.}
\indexauthor{Collaborator, A.}
\indexauthor{Author, L.}

\abstract{
The New Robotic Telescope (NRT) is an international collaboration to build and operate a 4 m
diameter fully robotic telescope. The telescope will take advantage of the superb atmospheric
conditions at the Observatory of the Roque de los Muchachos (ORM). In conjunction with a large
aperture, entirely robotic operation, quick response, and a set of versatile instrumentation in
the optical and near-infrared this guarantees a high scientific impact focused mainly in the area of
time domain astronomy. This paper presents the scientific motivation and the status of the
project, discussing possible technical solutions under evaluation for the optics, mechanics and
control system. }

\resumen{
El nuevo telescopio rob\'otico (NRT) es una colaboraci\'on internacional para construir y
operar un telescopio completamente rob\'otico de 4 m de di\'ametro. El telescopio aprovechar\'a
las excelentes condiciones atmosf\'ericas en el Observatorio del Roque de los Muchachos (ORM),
que junto a su gran apertura, su operaci\'on enteramente rob\'otica, su r\'apida  respuesta y
un conjunto de vers\'atil y variada instrumentaci\'on  en el \'optico e infrarrojo cercano
garantiza una contribuci\'on cient\'\i fica de alto impacto enfocada principalmente en la
novedosa \'area de la astronom\'\i a de dominio temporal. Este art\'\i culo presenta la
motivaci\'on cient\'\i fica  y el estado del proyecto, discutiendo posibles soluciones 
t\'ecnicas bajo evaluaci\'on para la \'optica, la mec\'anica y el sistema de control.}

\addkeyword{robotic telescopes}
\addkeyword{time-domain astronomy}

\begin{document}
\maketitle

\section{Introduction}

The New Robotic Telescope (NRT) is an international collaboration that aims to build and
operate an entirely robotic telescope with a diameter of 4 meter. The project started about
five years ago with Liverpool John Moores University (LJMU, UK) and the Instituto de
Astrof\'\i sica de Canarias (IAC, Spain) as promoters, and the Instituto Universitario de
Ciencias y Tecnolog\'\i as Espaciales de Asturias (ICTEA, Spain), the National Astronomical
Observatories,  Chinese Academy of Sciences, China (NAOC) and the National Astronomical
Research Institute of Thailand (NARIT) incorporated recently to the project.
Currently, the project is finishing its conceptual design to enter a more advanced design phase
with the aim to start operation in about five years time.

The motivation of the project is to develop an astronomical facility that will play a relevant
role in the forthcoming era of time domain astronomy by providing key complementary
observations for follow-up programmes in order to identify and characterize the most promising
targets discovered by other large facilities. Previous experience within the consortium 
building and operating the 2 m Liverpool robotic telescope \citep{steele04}, and developing
forefront astronomical instrumentation is a solid base for a proper execution of the project.
The strength of the NRT will rely on having the largest collector area for a telescope working in
robotic mode, a set of at least five instruments simultaneously installed in the different
focal stations, an efficient system to evaluate the priority of the targets and a quick
response that is essential for some of the scientific cases. 

The telescope will be sited at the Carlsberg Meridian site on the Observatorio del Roque de Los
Muchachos (ORM) in La Palma (Canary Islands, Spain) at 2,400 m altitude. Extensive site
testing campaigns for over thirty years have demonstrated \citep{wilson99}, that the ORM
has excellent 
conditions for astronomical observations, e. g. a mean seeing of 0.69 arcsec ($\sim 20$ \% of
the time below 0.5 arcsec) and an extremely dark sky ($\sim  22$ mag\, arcsec$^{-2}$ in V),
at the level of the best places in the world. The location allows  access to $\sim 3/4$ of
the sky, sharing $\sim 10,000$ square degrees with Southern hemisphere facilities.

This contribution presents a brief updated report on some of the most relevant analysis
conducted within the current conceptual design phase of the project. Previous reports of the
project have been presented by \citet{cop15} and \citet{gut19}. We refer readers to those
reports for further details of the project and in particular for topics like scientific cases,
site, dome concepts and instrumentation scarcely considered here. 

The paper is structured as follows: after this introduction, Section 2 enumerates some of the
most interesting scientific cases; Sections 3 and 4 are devoted to the optics and optomechanics
of the telescope, whilst Sections 5 and 6 sketch the overall concept of the mechanical structure
and the control system of the telescope. Section 7 presents the conclusions. 

\section{Scientific motivation}

The NRT will be very well suited in terms of collecting area (4$\pi$ square meters) and quick
response ($\sim 30$ sec on target) taking an intermediate position between the large ($> 8$ m
telescopes) and other  existing or planned fully robotic telescopes ($\sim 1-2$ m). Although
the NRT
has been conceived as a multipurpose telescope with instrumentation covering most of the
observing modes in the full optical and near-infrared ranges, the scientific motivation and
therefore most of the observing time will likely be dedicated to the area of time domain and
multimessenger astronomy. For instance, programmes to validate models of supernovae detected
by, e. g., the LSST \url{http://lsst.org}) or to measure
their light curves could be efficiently conducted by NRT. A quick response will also be essential
to observe the optical afterglows of gamma ray bursts (GRBs). Space missions like the Einstein
Probe \citep{yuan16} of the Chinese Academy of Sciences will discover various types of
high-energy transients in the soft X-ray band in large numbers, including tidal disruption
events (TDEs), supernova shock breakouts, high-redshift GRBs providing a huge number of targets
for NRT identification and characterization. The size and robotic operation of the NRT is also
ideal to study exoplanets by long follow-up programmes with appropriate instrumentation to
measure, for instance, the variations in the polarization of reflected light of exoplanets along
the orbital phase. 

The fully autonomous way of NRT operation will allow also to conduct a plethora of scientific
cases that require multiwavelength follow-up observations in coordination with other
facilities. Likely objects selected from the Gaia catalogs \citep{gaia16} will be one of the
main sources of targets.  

NRT observations will also be very relevant to identify and characterize the electromagnetic
counterparts of gravitational waves  \citep{abbot16}  and astronomical neutrinos \citep{ice18}. 

\section{Optical Design}

\label{sec:OptDesign}

Some of the main  NRT scientific cases related with transient objects (see above) require a
quick response. This has been taken into account in the optical and mechanical design, favoring
the selection of a compact structure that facilitates the telescope dynamics. However,
that concept increases the sensitivity of the system to mirror alignment, so a balance had to
be achieved between the optical and mechanical configurations. 

\begin{figure}[!t]
  \includegraphics[width=\columnwidth]{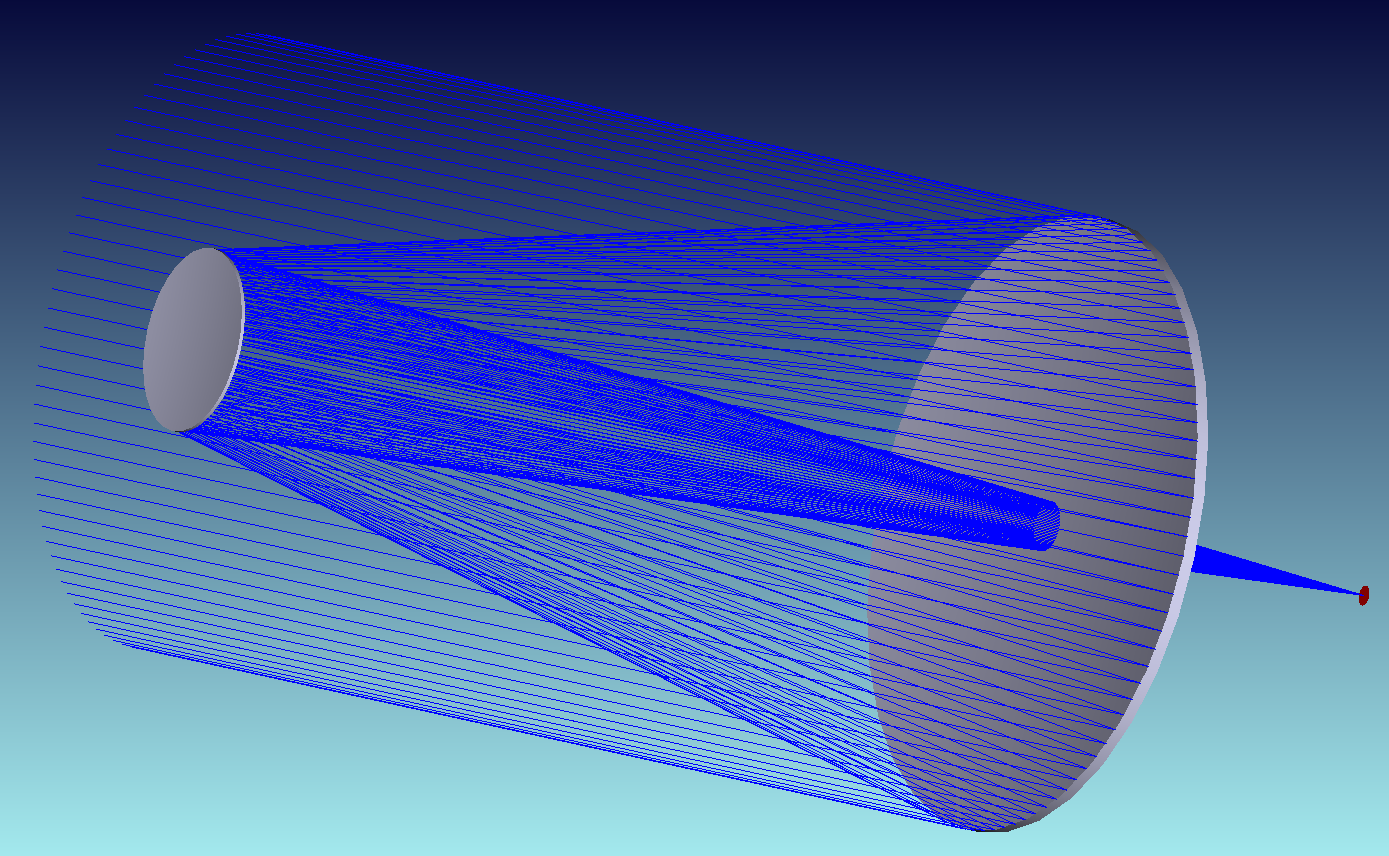}
  \caption{NRT optical design for a f/7.5 Ritchey-Chr\'etien configuration.}
  \label{fig:optlayout}
\end{figure}

Initially, a f/7.5 Ritchey-Chr\'etien \citep{1972Wetherell} configuration was considered, with
a fast f/1.75 primary mirror M1, i. e., a compact tube with a length of $\sim 7$ m.  A
representation of this  layout can be seen in  Figure~\ref{fig:optlayout}. Although this
configuration provides a very compact structure, it turned out to be very sensitive to alignment,
and the required mirror quality pushed the limits of the state of the art. That motivated us to
consider alternative designs  like the Dall-Kirkham \citep{2005Beach} with a spherical
secondary mirror M2. This concept provides a system which is an order of magnitude less
sensitive  to
mechanical deformations and thermal effects, and is comparatively much easier to collimate
and align. A spherical secondary mirror also reduces the complexity of manufacturing as
compared to the  hyperbolic mirror in the Ritchey-Chr\'etien design. The relatively worse
performance off-axis and the need for a more powerful corrector in the Dall-Kirkham optics as
compared to the Ritchey-Chr\'etien, may be assumed for a telescope like NRT that will be
dedicated mainly to spectroscopy. An ongoing in deep analysis of  performance,
sensitivity, manufacturing complexity and cost of both optical systems will help to reach a final decision. 

\subsection{Segmented primary mirror}
\label{sec:segprimary}

Monolithic primary mirrors are the common choice for 4 m-class telescopes, e. g.  the WHT \citep{2014WHT},
the SOAR \citep{2004SOAR}, and the DCT \citep{2018DCT}. However, segmented mirror topologies initially developed to overcome
the maximum practical size of monolithic mirrors (about 8 m), are starting to be
applied to smaller apertures, e. g. the LAMOST \citep{2010LAMOST} and  the SEIMEI \citep{2019SEIMEI} telescopes, due to a number of advantages in terms of manufacturing, cost, weight and recoating. In fact, a segmented approach allows to reduce the weight of the primary mirror  which is an important contributor to the total telescope mass. It also reduces the telescope downtime
associated to mirror coating, and facilitates all the mirror handling operations. Furthermore, there is
 a broader choice of mirror manufacturing options for smaller mirrors, and a
segmented design is scalable to larger apertures. 

\begin{figure}[!t]
  \includegraphics[width=\columnwidth]{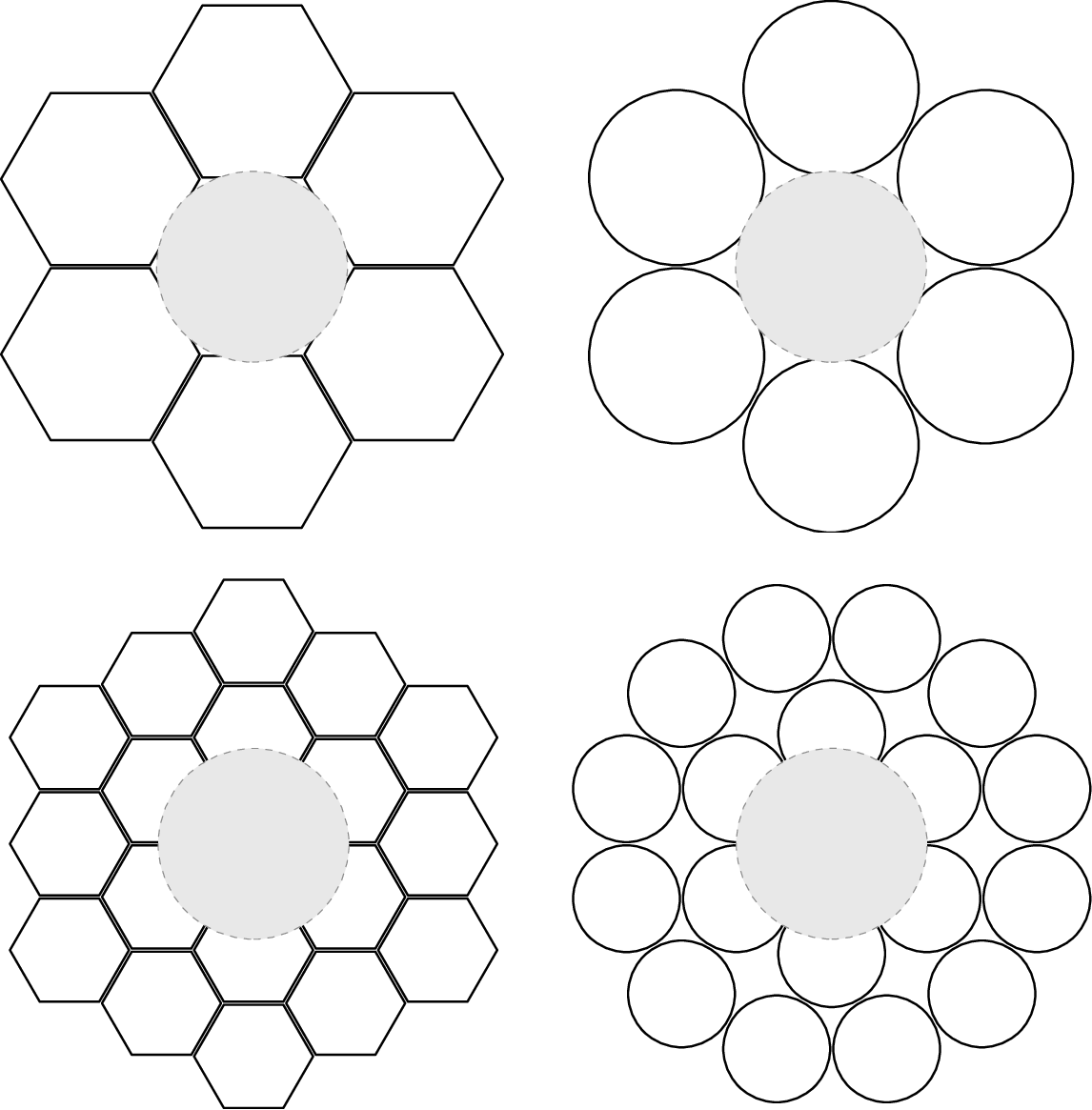}
  \caption{NRT primary mirror segmented topologies, for 6 and 18 hexagonal and circular segments. The shaded region in the center of each configuration corresponds to the area obscured by the secondary mirror.}
  \label{fig:topologies}
\end{figure}

Several options for   the segmented topology of the primary mirror of NRT are under study and
evaluation. Some of them, based on arrangements of hexagonal and circular mirrors, are shown  in
Figure~\ref{fig:topologies}. The performance of the telescope and the sensitivity to segment
alignment are being evaluated for each topology. In general, hexagonal segments allow for more
compact apertures, but present more difficulties to manufacture. Circular segments are easier
to manufacture but the optical performance of the telescope is worse in the near infrared.
Other considerations affecting each topology are considered in next sections.

\section{Optomechanics}
\label{sec:Optomechanics}

This section presents the different alternatives that are under consideration for the optomechanics of the primary and secondary  mirror assemblies of the telescope.

\subsection{Mirror Support System}

The deformation of the reflecting surfaces of the mirrors has to be controlled to meet the telescope optical error budget. That requires an adequate
selection of the substrate material (Zerodur in our case), the mirror geometry and the 
support system in order to avoid transmitting the deformations produced from the
mirror interfaces to the optical surface.

Pointing to different altitudes requires a mirror support
system able to sustain the mirror weight in both the axial and lateral
directions at any altitude. The challenges on how to support gravity are very different in those
two directions, and for this reason the support system of the mirrors is separated
into two almost independent components, each of them stiff in one direction and compliant in the other. The baseline design for
both the M1 segments and the M2 mirror is the general solution adopted in other segmented telescopes with a mechanical whiffletree for axial support
and a central diaphragm for lateral support.

\begin{figure}[!t]
  \includegraphics[width=\columnwidth]{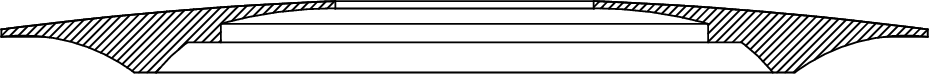}
  \caption{Section view of the M2 substrate, showing the double-arch lightweighting solution.}
  \label{fig:M2Baseline}
\end{figure}

In parallel to this analysis, a preliminary study of different lightweighting options for the
M2 mirror has been performed. The objective of a lightweighted M2 is to reduce the overall tube
weight and to match the hexapod specifications to an available commercial solution. In the
scope of this first study, simulations to optimize the geometry of the M2 substrate minimizing
the axial gravity print-through and the mirror mass have been conducted. The results of the
simulations have led to the concept presented in Figure~\ref{fig:M2Baseline}. This M2 design is
based on a double arch substrate with a reduced weight of about 100 kg, which represents a
lightweighting factor of $\sim 50$ \%. 

\subsection{Active Optics}

Active optics is essential to relax the requirements of the telescope structural deformations, therefore allowing for a lighter structure. Such active corrections will be incorporated into 
NRT by M1 and M2 mirror repositioning. In the case of a segmented configuration, The M1 active
optics system will be comprised by three linear actuators per segment operating in a closed
loop with a set of sensors that guarantee the relative position between mirrors and
keep the global shape of the M1 optical surface. The alignment between the M1 assembly and the M2 mirror will be actively controlled with the M2 hexapod.

Topologies with one ring of six circular or hexagonal segments (see Fig. 2) have less system complexity and easier
active optics control as compared to the 18 segments ones, but obviously they require bigger and heavier segments. On the other hand, topologies with a higher number of segmented mirrors have more correction capabilities to the overall primary mirror shape and they also have broader options for COTS (commercial of the shelf) segment actuators due
to the smaller size and weight. If COTS
solutions are finally adopted, the efforts will be directed to a proper  definition of the specifications of the systems during the
telescope preliminary design.

\section{Telescope Structure}

The elevation structure will be an open
truss with two different variants under consideration: a classical Serrurier and a
tripod-like structure. As discussed in the optomechanics section, active optics
will be implemented for the telescope so the ability to maintain
collimation by passive means is not mandatory and then the tube deformation requirements
can be relaxed. The classical Serrurier has the benefit of reducing the M1
obscuration, however the tripod is structurally advantageous, lighter and the obscuration
problem can be limited if the tube beam locations are matched with the spaces naturally present
in a segmented telescope. A light upper tube (with a lightweighted M2 and a tripod-like tube) moves the M1 assembly closer to the elevation axis to balance the tube center of gravity  and permits a direct attachment to the elevation ring.
Three proposals are being considered for the azimuth
mount: conventional yoke (baseline), gantry
and rocking chair.

\begin{figure}[!t]
  \includegraphics[width=\columnwidth]{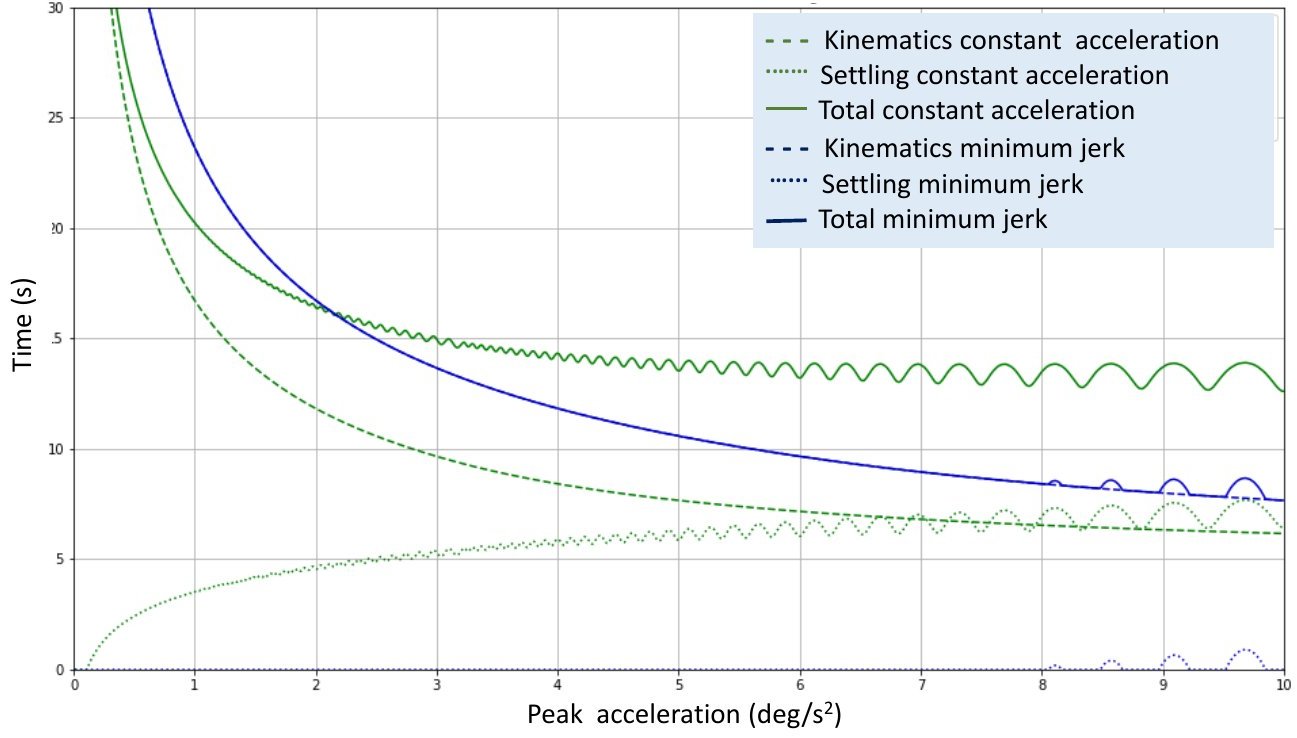}
  \caption{Blind pointing time vs peak acceleration for a 70 degrees slew, comparing the trapezoidal (green) and the S-curve (blue) profiles.  }
  \label{fig:BlindPointingTime}
\end{figure}

The telescope structure design has to include the slewing time assessment as quick response is
one of the key features of the telescope. A preliminary model has confirmed that the slew
requirements can be fulfilled with commercially available solutions. That  model considers both
the slew kinematics and the associated settling time assuming a very simple single degree of
freedom model subjected to viscous damping. Two motion profiles have been evaluated: a constant
acceleration profile (trapezoidal in velocity) and an S-curve profile of minimum jerk that
limits the settling times. Figure~\ref{fig:BlindPointingTime} shows a comparison of the time
required to slew and settle the tube position for both profiles and different peak
accelerations. The dynamic inverse problem has also been solved for each axis to compute the
required driving torques, predict slew time for a given motor, and in this way establish an
admissible maximum length for the telescope tube. A complete trade-off of the different
configurations is still to be performed.

\section{Control System}
\label{sec:ctrlsystem}

The telescope control system (TCS) is a key concept in the design of NRT. The TCS will be
responsible for controlling, coordinating, monitoring and planning all systems (hardware and
software) involved in the operation of the telescope, it must guarantee the proper operation of
all systems and services and the security in terms of fault tolerance, problem recovery and
safe shutdown in the event of a serious failure. A fully autonomous operation requires a
constant exchange of information between the different subsystems and the TCS in order to
determine and execute at any moment the most adequated actions. This requires quick responses 
and to establish a set of priorities according to the state of the telescope and its
environment, e. g., atmospheric conditions at any moment.

The NRT control system architecture aims to follow best practices in services decoupling and
deployment, following recent techniques in containerization and orchestration. This type of
system will give a great stability, scalability, and flexibility, allowing new services to be
added or removed without affecting the operation of the others.

The current robotic telescopes in operation have simple TCSs as compared to the requirements
of the NRT. Therefore, in this phase of conceptual design we are evaluating pros, cons, suitability
and efforts needed to adopt a new solution.  There are several platforms, such as ROS (Robotic
Operating System, \url{https://www.ros.org/}), INDI (Instrument Neutral Distributed Interface,
\url{https://www.indilib.org/}), ACS (\url{http://www.eso.org/~almamgr/AlmaAcs/index.html}) or
TANGO (\url{https://www.tango-controls.org/}), which offer a base to develop a TCS. However,
independently of their relative value and suitability for NRT, all of them require a high
effort  in implementation, adaptation and verification. As a promising alternative to those
systems, we are exploring the possibility to adapt GCS, the control system of GTC \citep
{1998Filgueira}, running successfully in that telescope and having a development team with many
years of experience.  The paragraphs below describe the main facts of GCS  and the inherent
problems we face to adapt it to NRT.

GCS has advantages like the execution of processes in kernels with real time, a simulator and a
development platform, as well as a good log record to detect and correct problems.  However, it
also has  some disadvantages with respect to other platforms: it uses CORBA (Common Object
Request Broker Architecture, \url{https://www.corba.org/system)} which is a bit outdated and
perhaps GCS is too complex for the NRT requirements. An implementation using DDS (Data
Distribution Service, \url{https://www.dds-foundation.org/}) might be more convenient.

GCS is a complex system, and has thus been decomposed into different sets of subsystems that are
formed by groups of devices. These subsystems are responsible for the operation of the
telescope performing the tasks assigned to them: Logging \& Alarms Service, Observing Engine,
System Monitor, Data Factory, Sequencer, Scheduler, Acquisition \& Guiding Control System
(AGCS), Science Instrument Control System (SICS), Enclosure and Services Control System (ESCS),
Configuration Service, Primary Mirror Control System (M1CS), Interlock \& Safety Control System
(ISS), Main Axis Control System (MACS), etc.

The communications between objects are via a network using an implementation of the CORBA
standard for these devices. Within this model, a client device can invoke a method exported by
another server device with location transparency. This structure allows for a great robustness,
flexibility and fault tolerance.  The devices are defined using a plain text file, from which
the development tools generate a skeleton in C++ or Java. The generation of this code skeleton
is done with a command line tool. The user-written source code may be regenerated multiple
times without affecting previous compilations. As in object-oriented programming, devices can
inherit from other devices, and there is also the possibility of grouping several devices to
create a more complex structure. The devices inherit the CORBA communication interfaces and
methods of the base class, which allows them to communicate with the other services and
subsystems of the control system. This feature allows the devices to interact with the rest of
the system to connect to the databases, obtain information on the status of the telescope, be
aware of alarms, and allow the control system to know the status of the device at any time.

GCS currently does not implement any subsystem for robotic control, since the GTC works in the
traditional way with human operation. The challenge of adapting GCS to the NRT control system
will be the integration of the robotic control of the telescope, allowing a completely
autonomous operation.

\begin{figure}[!t]
  \includegraphics[width=\columnwidth]{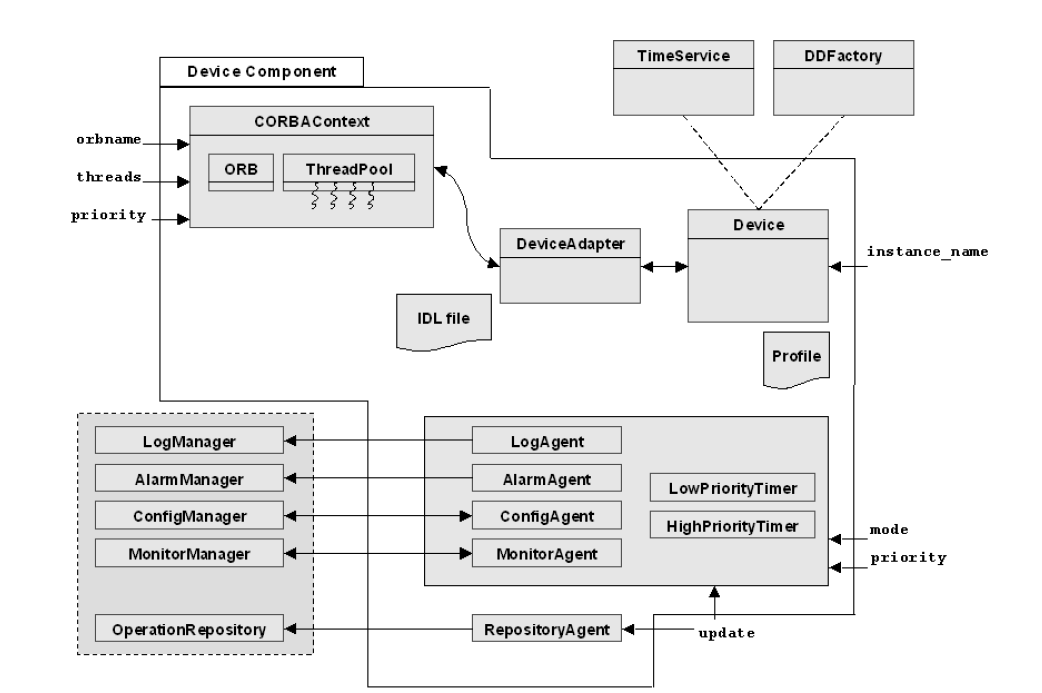}
  \caption{GCS device component model \citep{2012Rodriguez}. }
  \label{fig:GCS}
\end{figure}

The NRT aims to keep the GCS model of decoupling system components, having distributed
execution and communications, but migrating the CORBA middleware to a more modern approach,
like DDS. Another advantage is the abstraction from low-level hardware and software GCS offers
when integrating new devices into the system, since all new devices start from a common scheme
or class-base. A tight collaboration between the GTC and NRT software teams has been
established to identify the challenges, advantages and drawbacks to adapt and extend such a
system to the requirements and operation of a robotic telescope like the NRT. A good understanding
of GCS and the other alternatives will help to design the architecture of the future NRT
control system.

\section{Conclusions}

This contribution has presented an updated report of the NRT project focused on the status of
the optical, mechanical and control system of the telescope design. NRT will be the first 4 m
telescope with fully robotic operation in the world.  It will  play an essential role within
the context of the large astronomical ground-based and space facilities that are planned
the coming years.  When it enters into operation in about five years time, NRT will be a world-leading telescope for quick identification and characterization of the most interesting
transients discovered by such facilities. An international consortium formed by several
institutions from four countries has been developed and we created a strong scientific and
technical team that is identifying and planning programmes to accomplish the most ambitious
scientific cases and defining the best technical solutions for the telescope. The NRT will
also be
useful to validate concepts and operations with the aim to establish standards for future
generations of larger and larger robotic telescopes. 

\section*{Acknowledgements}
This Project has received funding from the Canary Islands Government under a direct grant awarded on grounds of public interest (Order Nº 185 August 2017).

\end{document}